\documentclass[12pt]{article}

\usepackage{scicite}
\usepackage[T1]{fontenc}

\usepackage{graphicx,graphics,amsfonts,amsmath,amssymb}

\usepackage{times}
\usepackage{graphicx}

\usepackage{color}
\usepackage{verbatim}
\usepackage{tikz}

\makeatletter
\renewcommand*{\@fnsymbol}[1]{\ensuremath{\ifcase#1\or \dagger\or \ddagger\or
    \mathsection\or \mathparagraph\or \|\or **\or \dagger\dagger
    \or \ddagger\ddagger \else\@ctrerr\fi}}
\makeatother

\newenvironment{sciabstract}{%
\begin{quote} \bf}
{\end{quote}}

\newcommand{\beginsupplement}{%
        \setcounter{table}{0}
        \renewcommand{\thetable}{S\arabic{table}}%
        \setcounter{figure}{0}
        \renewcommand{\thefigure}{S\arabic{figure}}%
        }

\newcounter{lastnote}

\title{Kibble-Zurek universality in a strongly interacting Fermi superfluid}

\author
{Bumsuk Ko$^{1, 2}$, Jee Woo Park$^{1\ast}$, and Y. Shin$^{1, 2\ast}$\\
\\
\normalsize{$^{1}$Department of Physics and Astronomy, and Institute of Applied Physics,}\\
\normalsize{Seoul National University,}\\
\normalsize{Seoul 08826, Korea}\\
\normalsize{$^{2}$Center for Correlated Electron Systems, Institute for Basic Science,}\\
\normalsize{Seoul 08826, Korea}\\
\\
\normalsize{$^{\ast}$To whom correspondence should be addressed;}\\
\normalsize{E-mail: jw\_park@snu.ac.kr, yishin@snu.ac.kr.}\\
}

\date{}

\begin{document}


\baselineskip24pt


\maketitle


\begin{sciabstract}
Near a continuous phase transition, systems with different microscopic origins display universal dynamics if their underlying symmetries are compatible. In a thermally quenched system, the Kibble-Zurek mechanism for the creation of topological defects unveils this universality through a characteristic power-law exponent, which captures the dependence of the defect density on the quench rate. Here, we report the observation of the Kibble-Zurek universality in a strongly interacting Fermi superfluid. As the system's microscopic description is tuned from bosonic to fermionic, the quench formation of vortices reveals a constant scaling exponent arising from the $U(1)$ gauge symmetry of the system. For rapid quenches, destructive vortex collisions lead to the saturation of their densities, whose values can be universally scaled by the interaction-dependent area of the vortex cores.
\end{sciabstract}


The central premise in the study of continuous phase transitions is the notion of scale invariance and universality. In the vicinity of a critical point, the characteristic length and time scales of a system's correlations diverge, giving rise to collective phenomena that are largely decoupled from the microscopic details of the system. The Kibble-Zurek (KZ) mechanism for the creation of topological defects presents a paradigmatic example of such universal dynamics~\cite{Kibble1976, Zurek1985}. In a system whose temperature $T$ is cooled across the critical temperature $T_{\rm c}$ for a continuous phase transition, its diverging relaxation time $\tau$ will eventually exceed the timescale for the change of temperature, at which point the system can no longer follow the externally imposed quench. Consequently, the system's correlation length $\xi$ becomes frozen at a value $\hat{\xi}$ until its reflexes are recovered in the broken-symmetry phase. This leads to the formation of domains of size $\hat{\xi}$ with independent order parameters, which merge to create topological defects at their boundaries. The essential prediction of the KZ theory is the existence of a universal scaling relationship between the defect density and the quench rate. Specifically, for a linear temperature quench, the defect density will have a power law dependence on the quench rate, with a characteristic exponent $\alpha_{\rm KZ}$ determined by the critical exponents $\nu$ and $z$ of the system. Here, $\nu$ and $z$ describe the divergence of $\xi$ and $\tau$ through $\xi = \xi_{0} |(T-T_{\rm c})/T_{\rm c}|^{-\nu}$ and $\tau = \tau_{0} |(T-T_{\rm c})/T_{\rm c}|^{-\nu z}$, where $\xi_{0}$ and $\tau_{0}$ are the microscopic length and time scales of the system, respectively. Since systems that fall into the same universality class have a shared set of critical exponents determined by generic factors such as the underlying symmetries, range of interactions, and dimensionality, the KZ exponent will also be universally determined by these factors. The KZ mechanism has been explored in a number of systems including superfluid He~\cite{Hendry1994, Bauerle1996, Ruutu1996}, liquid crystals~\cite{Chuang1991}, ion chains~\cite{Pyka2013, Ulm2013, Ejtemaee2013}, and atomic Bose-Einstein condensates~\cite{Weiler2008, Lamporesi2013, Navon2015}, but a systematic exploration of its universal nature within a single physical system has yet been demonstrated.

Strongly interacting atomic Fermi gases with tunable interactions offer a unique opportunity to explore the KZ mechanism in a setting where the system's microscopic description can be consistently tuned. In the vicinity of a Feshbach resonance, as the scattering length between the spin-up and spin-down fermions is tuned from positive to negative, the nature of superfluidity changes from Bose-Einstein condensation (BEC) of tightly bound molecules to Bardeen-Cooper-Schrieffer (BCS) superfluidity of long-range fermion pairs~\cite{Giorgini2008, Zwerger2011}. This leads to significant variations in the static and dynamic properties of the superfluid, as apparent in the changes of the critical temperature~\cite{Nozieres1985, Regal2004, Chen2006} and the critical velocity~\cite{Miller2007, Weimer2015, Park2018} in the BEC-BCS crossover. Nevertheless, the spontaneous breaking of the $U(1)$ gauge symmetry associated with the normal to superfluid phase transition should prevail throughout the crossover, and the question of whether if the KZ mechanism can be consistently applied in this regime remains largely unaddressed.

Here, we report the observation of the Kibble-Zurek universality in a strongly interacting Fermi superfluid. A linear temperature quench in an oblate sample of $^{6}$Li atoms near a Feshbach resonance spontaneously creates as many as 50 vortices in the superfluid phase, whose counting statistics reveals the characteristic power-law scaling of the KZ mechanism. As the interaction is tuned across the BEC-BCS crossover, the extracted critical exponents remain constant at a value that is consistent with the predictions of the inhomogeneous KZ mechanism for a BEC in a harmonic trap~\cite{delCampo2011}. Interestingly, when the quench rate is sufficiently increased, the vortex density becomes saturated to a value that is strongly dependent on the interaction strength. This is explained in terms of the destructive collisions among vortices with opposite sign, whose effective range is governed by the interaction dependent vortex core size of the superfluid. This mechanism reveals the microscopic scales involved in the critical dynamics of the superfluid, allowing to collapse the data obtained in different regimes of the BEC-BCS crossover to a single universal curve.

\begin{figure*}[p]
\centering
    \includegraphics[width=1\columnwidth]{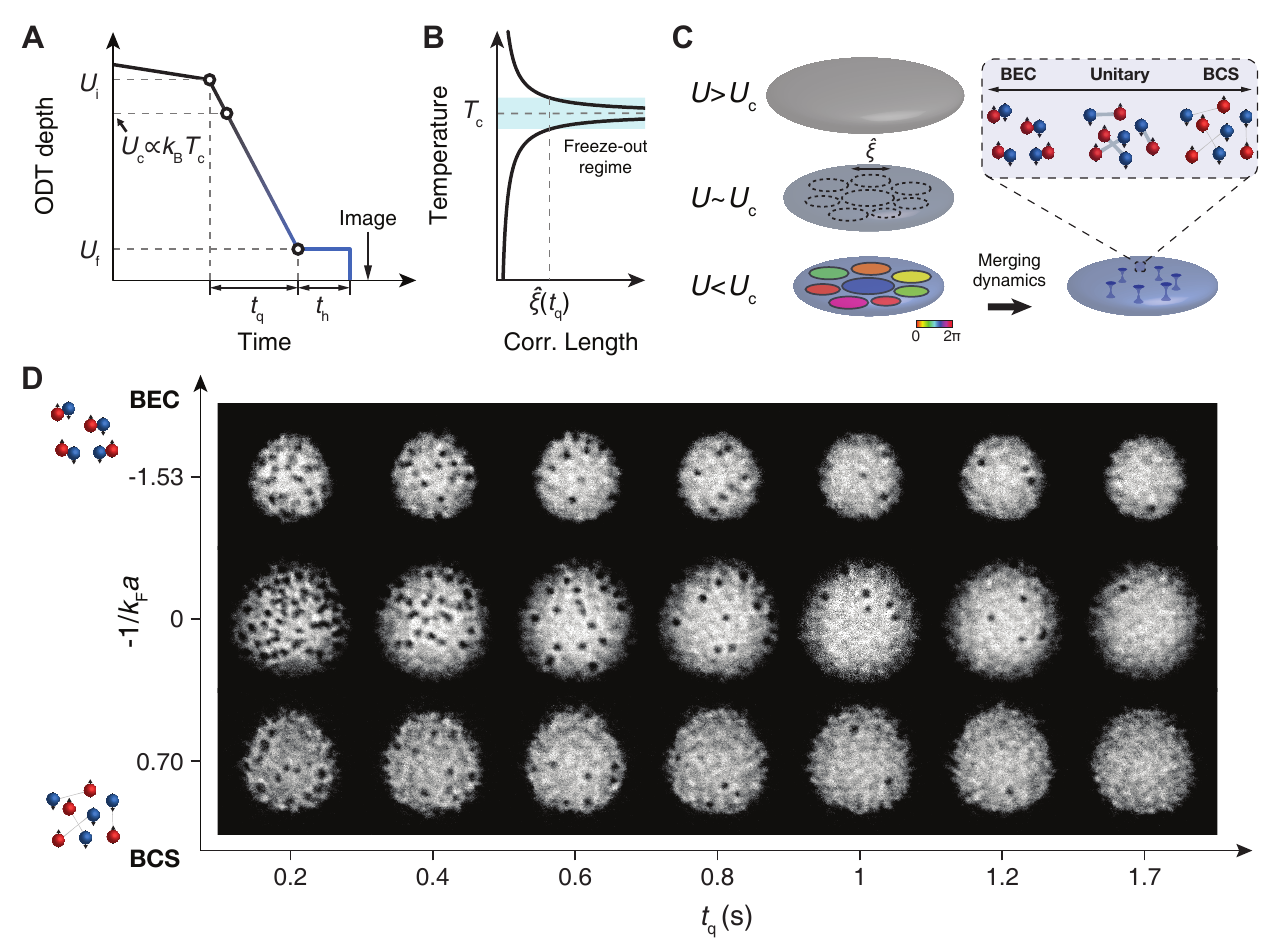}
  \caption{\textbf{Spontaneous defect formation during the normal to superfluid phase transition in a strongly interacting Fermi gas.} ({\bf A}) Schematic of the experiment. The ODT depth is initially set to $U_{\rm i} = 1.15~U_{\rm c}$, and the gas is evaporatively cooled by linearly lowering the trap depth to $U_{\rm f}$ in a variable duration $t_{\rm q}$. After a hold time $t_{\rm h}$, the sample is released from the trap and an absorption image of the sample is taken. ({\bf B}) Evolution of the correlation length and its freeze-out during the phase transition. During the quench, the critical slowing down of the system's reflexes near $T_{\rm c}$ leads to the freeze-out of the diverging correlation length at a value $\hat{\xi}$. In the experiment, since the quench width $U_{\rm i}-U_{\rm f}$ is fixed, the quench time $t_{\rm q}$ determines the value of $\hat{\xi}$. ({\bf C}) Formation of quantized vortices during spontaneous symmetry breaking. Upon entering the broken-symmetry phase, superfluid domains with independent phases are formed, which merge to create quantized vortices at their boundaries. Patches with different colors indicate domains with different phases. The defect formation dynamics is explored across the BEC-BCS crossover. ({\bf D}) Exemplary images of the gas after time-of-flight. The number of created vortices are recorded as a function of the quench duration and the interaction strength.
}
\label{Fig1}
\vspace{3cm}
\end{figure*}

The experiment starts with the creation of an equal mixture of ultracold $^{6}$Li in their two lowest hyperfine states $|1\rangle$ and $|2\rangle$ near the broad $s$-wave Feshbach resonance located at 832 G~\cite{Park2018, supp}. The atoms are confined in a highly oblate trap that consists of a single beam optical dipole trap (ODT) providing the tight axial confinement and a weak magnetic trap providing the radially symmetric confinement. Initially, about $2.0 \times 10^{6}$ atoms per spin state are prepared at different magnetic fields around the Feshbach resonance, at a temperature corresponding to 1.15 times the critical ODT depth $U_{\rm c}$ for the onset of condensation. Then, the mixture is evaporatively cooled by linearly reducing the ODT depth to a final value $U_{\rm f}$ during various quench times $t_{\rm q}$ ranging between 0.2 s and 2.6 s [see Fig.~\ref{Fig1}A]. In the experiment, $U_{\rm f}$ is set to $0.15~U_{\rm c}$ for all the measurements except for those performed on the BCS side of the resonance, where $U_{\rm f}$ is set to $0.3~U_{\rm c}$ to maintain a moderate condensate size. The linear relationship between the sample temperature and the ODT depth is checked by monitoring the momentum width of the thermal clouds, and a study of the condensate growth dynamics in the experiment shows that the system is well thermalized for the explored range of quench times~\cite{supp}.

During the quench, the system undergoes a spontaneous symmetry-breaking phase transition, and superfluid domains with independent phases are formed [see Fig.~\ref{Fig1}B, C]. Since the ODT provides the tight axial confinement, lowering the ODT depth leads to the evaporation dominantly in the axial direction along which fast thermalization is ensured. This leads to the production of domains principally in the radial plane of the sample, which merge to create vortices aligned parallel to the axial direction of the trap. To facilitate the merging dynamics, a hold time $t_{\rm h}$ of 200 ms (50 ms for the measurements on the BCS side) is applied after the quench, and finally, the vortices are detected by performing time-of-flight imaging of the sample using resonant light~\cite{Park2018}. At the end of this procedure, the sample contains about $1.0 \times 10^{6}$ atoms per spin state with a condensate fraction of 80$\%$ at unitarity. The final radial trap frequency is 17 Hz, and the axial trap frequency ranges between 350 Hz and 400 Hz, depending on the explored interaction regimes. Representative images of the sample for a range of quench times are shown in Fig.~\ref{Fig1}D for different interaction strengths $1/k_{\rm F}a$. Note that the values of $1/k_{\rm F}a$ represent the final state of the gas, where $k_{\rm F}$ is the Fermi wave number of non-interacting fermions in a harmonic trap and $a$ is the $s$-wave scattering length.

\begin{figure*}[p]
\vspace{1cm}
\centering
  \includegraphics[width=1\columnwidth]{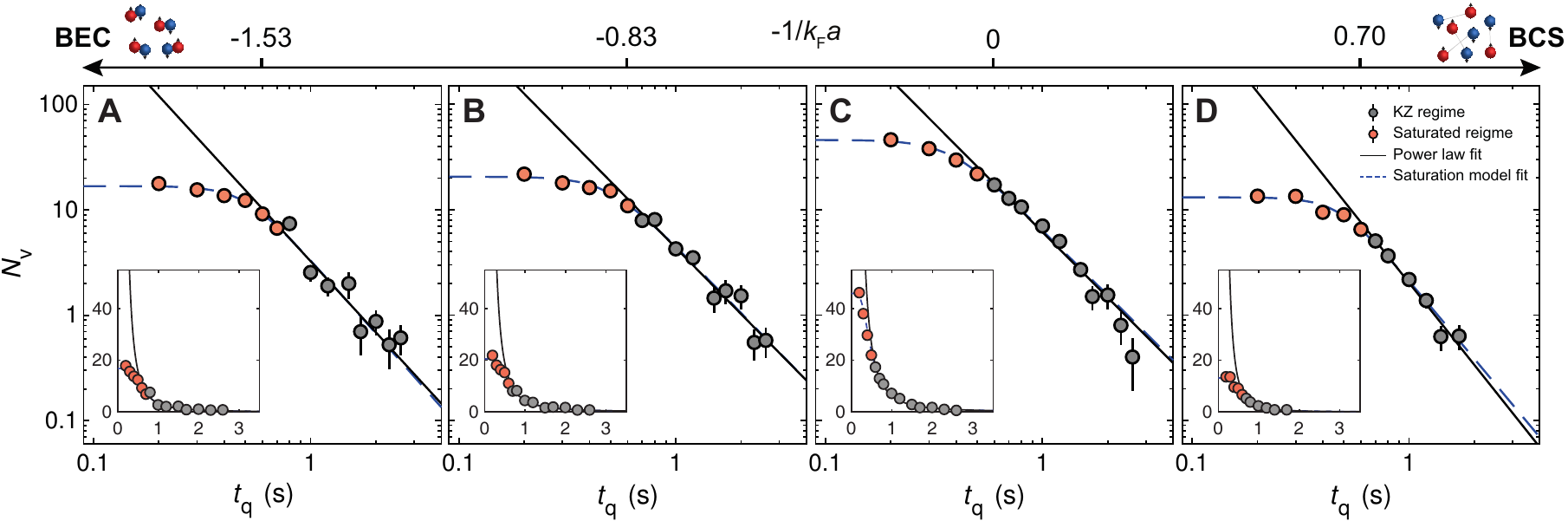}
   \caption{\textbf{Vortex number versus quench time.} The average number of detected vortices are plotted as a function of $t_{\rm q}$ in log-log scale at four different final interaction strengths $-1/k_{\rm F}a=$ ({\bf A}) $-1.53$ (757 G), ({\bf B}) $-0.83$ (785 G), ({\bf C}) 0 (832 G), and ({\bf D}) 0.70 (898 G). The gray and the orange circles represent the data points in the KZ and the saturated regime, respectively. The blue dashed line is the saturation model fit to the full regime, and the black solid line is the power law fit to the KZ regime. The insets show the same data in lin-lin scale. Each data point comprises at least ten realizations of the same experiment. The error bars are standard error of the mean.
}
\label{Fig2}
\vspace{3cm}
\end{figure*}

Fig.~\ref{Fig2} shows the average number of observed vortices $N_{\rm v}$ as a function of the quench time $t_{\rm q}$ at four different interaction strengths across the BEC-BCS crossover. Our measurements feature an order of magnitude higher number of defects (up to 50 at unitarity) compared to previous KZ experiments with atomic BECs, which results from the large radial size and the short healing length of the strongly interacting Fermi superfluid. The dynamic range of the defect number reaches about two orders of magnitude, and the quench duration spans over a decade, allowing different regimes of the vortex formation dynamics to be clearly distinguished in the experiment. Specifically, for sufficiently long quench times, the evolution of the vortex number reveals the characteristic power-law scaling of the KZ mechanism, but as $t_{\rm q}$ is reduced, the vortex number becomes saturated. Such behavior has been observed in previous KZ experiments~\cite{Ulm2013, Donadello2016, Liu2018}, and was attributed to the presence of destructive interactions among defects. To systematically distinguish the two regimes and reliably extract the KZ exponent, we adopt an empirical model of the defect number $N_{\rm v}=N_{\rm sat}[1+(t_{\rm q}/t_{\rm sat})^{2\beta_{\rm KZ}}]^{-1/2}$ proposed in Ref.~\cite{Donadello2016}, where $N_{\rm sat}$ is the saturated vortex number, and $t_{\rm sat}$ is the characteristic quench time for the onset of saturation. The model fits the data well in all interaction regimes, suggesting the existence of a universal scaling of the vortex number even in the presence of saturation. From these fits, we identify the KZ regime through the condition $t_{\rm q}\gtrsim1.5~t_{\rm sat}$, and a power-law function is fit in this regime to extract the KZ exponent $\alpha_{\rm KZ}$.

\begin{figure*}[p]
\vspace{1cm}
 \centering
  \includegraphics[width=0.45\columnwidth]{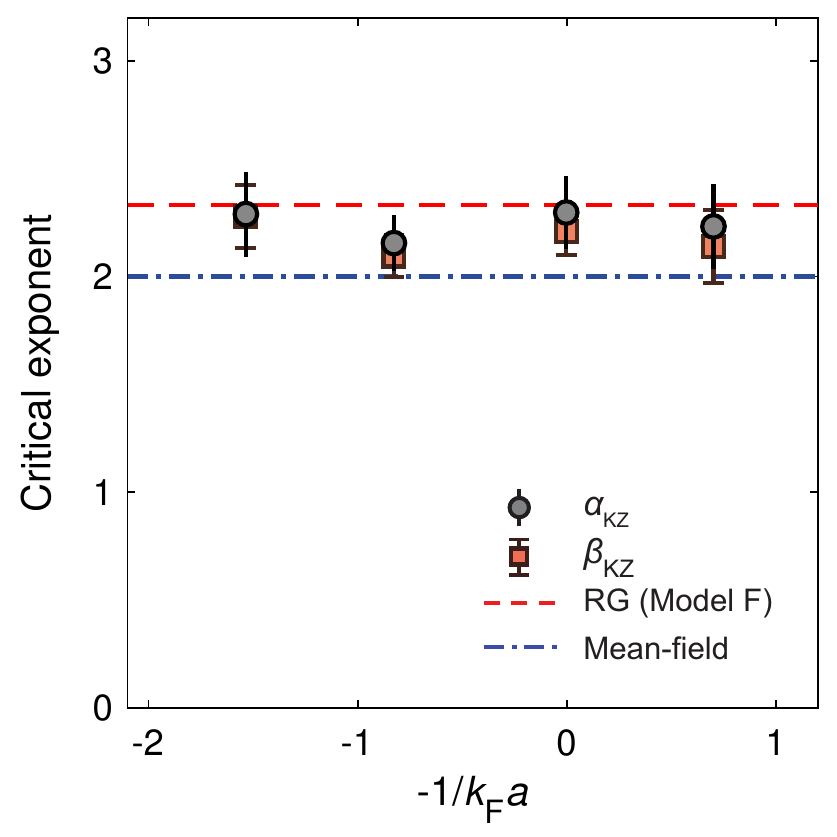}
  \caption{\textbf{Kibble-Zurek exponents in the BEC-BCS crossover.} The KZ exponents obtained from the power law fits to the KZ regime, $\alpha_{\rm KZ}$ (gray circles), and from the model fits to the full regime, $\beta_{\rm KZ}$ (orange squares), are shown for the explored interaction strengths. Each data point is the average of three sets of experimental measurements, and the error bars are the quadrature sums of the standard error of the mean and the fit error. The red dashed and the blue dot-dashed line indicate the prediction of the F-model and the mean-field, respectively. The values of $\alpha_{\rm KZ}$ and $\beta_{\rm KZ}$ at each each interaction strengths lie close to each other, justifying the condition for setting the KZ regime in the experiment.
}
\label{Fig3}
\vspace{3cm}
\end{figure*}

A summary of the measured KZ exponents for the experimentally explored interaction strengths are shown in Fig.~\ref{Fig3}. The central observation is that, within the experimental resolution, $\alpha_{\rm KZ}$ maintains a constant value across the BEC-BCS crossover, arising from the global $U(1)$ gauge symmetry of the system. The values of $\alpha_{\rm KZ}$ range between 2.15 and 2.30, and their average yields $\alpha_{\rm KZ}=2.24 (9)$. To put these numbers in perspective, we compare the measured values of $\alpha_{\rm KZ}$ to theoretical predictions of the KZ exponent in various scenarios. First of all, for a homogeneous system, the KZ exponent is given by $(D-d)\nu/(1+\nu z)$~\cite{Zurek1985}, where $D$ is the dimensionality of the system and $d$ of the defects. However, for an inhomogeneous sample confined in a harmonic trap, condensates with locally chosen phases are created only where the velocity of the transition front is faster than the propagation speed of the phase information, and a modification of the KZ exponent to 
\begin{align}
(D-d)\frac{1+2\nu}{1+\nu z}
\end{align}
has been predicted in Ref.~\cite{delCampo2011}. With $D-d=2$ for vortices created in a 3D sample and the values of $\nu$ and $z$ for the lambda transition of superfluid helium-4, the mean-field and the renormalization group theory (F-model) predictions of the KZ exponents are 1/2 and 2/3 for a homogeneous system and 2 and 7/3 for a harmonically trapped system, respectively. The measured $\alpha_{\rm KZ}$ departs far from the homogeneous case but lies between the predictions for a harmonic trap, closer to that of the F-model. However, an improved signal-to-noise is necessary to firmly distinguish between the mean-field and F-model predictions. In light of the Ginzburg criterion, our sample is expected to be away from the mean-field regime since its inter-particle spacing $1/k_{\rm F}\approx0.3~\mu$m, which sets the shortest length scale of the system, is larger than the characteristic Ginzburg length $\xi_{\rm G} = \lambda_{\rm dB, c}^2/(\sqrt{128}\pi^2|a|) \lesssim 0.15~\mu$m~\cite{donner2007}. Here, $\lambda_{\rm dB, c}$ is the thermal de-Broglie wavelength at the critical point.

The saturation of the vortex number at short quench times opens a window to explore the microscopic aspects of defect formation dynamics near the critical point. As mentioned earlier, the likely cause for this behavior is the destructive collisions among vortices and antivortices, which are consistent with the observed enhancement of the vortex decay rate in the saturated regime~\cite{supp}. To estimate the saturated vortex number and understand its dependence on $1/k_{\rm F}a$, we adopt a simple picture for the lossy collisions, where vortex annihilation is strongly enhanced when the inter-vortex distance becomes shorter than a characteristic distance set by the vortex core size $\xi_{\rm vor}$. Then, the maximal vortex number will be given by $(R_{\rm TF}/f \xi_{\rm vor})^2$, where $R_{\rm TF}$ is the Thomas-Fermi radius of the superfluid, and $f$ is a scaling factor that accounts for the characteristic distance. Since $f$ is a geometric factor that captures the effective size of a vortex, it is reasonable to expect that its value is fairly constant within the crossover. A mean-field estimate of $N_{\rm sat}$ can then be obtained by substituting $\xi_{\rm vor}$ with the superfluid healing length $\xi_{\rm h} = \hbar/mv_{\rm c}$, where $m$ is the atomic mass of $^{6}$Li and $v_{\rm c}$ is the mean-field Landau critical velocity of the superfluid sample at its center. The values of $R_{\rm TF}$ are experimentally determined from the images of the sample~\cite{supp}.

\begin{figure*}[p]
\vspace{1cm}
 \centering
  \includegraphics[width=0.85\columnwidth]{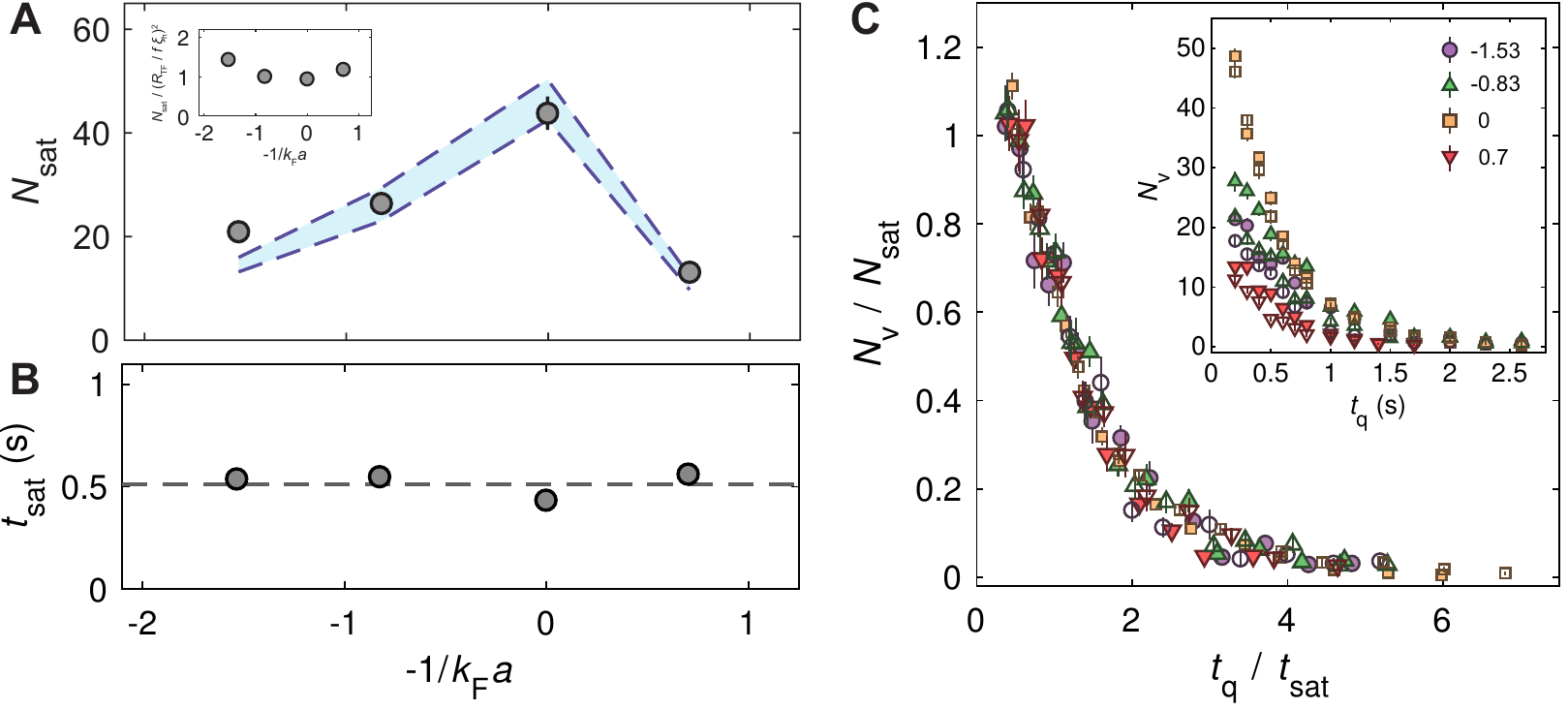}
  \caption{\textbf{Characterization of spontaneous vortex formation across the BEC-BCS crossover.} ({\bf A}) Comparison between $N_{\rm sat}$ (gray circles) and $R_{\rm TF}^2/(f\xi_{\rm h})^2$ (blue band), including a scaling factor of $f\approx40$. The inset shows the ratio between the two values. The data points are obtained from the model fit applied to a single set of experiments, and the error bars represent the corresponding fit error. The band comes from the uncertainties in $R_{\rm TF}$ and $k_{\rm F}$ in the experiment. ({\bf B}) Quench time $t_{\rm sat}$ at the onset of saturation. The error bars, which represent the corresponding fit error, are smaller than the data points. The dashed line is the guide to the eyes, indicating the average of the data points. ({\bf C})  Normalized vortex number $N_{\rm v}/N_{\rm sat}$ as a function of the normalized quench time $t_{\rm q}/t_{\rm sat}$ at four different interaction strengths $-1/k_{\rm F}a{=}-1.53$ (purple circles), $-0.83$ (green triangles), 0 (orange squares), and 0.7 (red inverted triangles) across the BEC-BCS crossover. The open markers show an independent measurement with slightly lower atom numbers. The error bars are standard error of the mean. The inset shows the original $N_{\rm v}$ versus $t_{\rm q}$ for comparison.
}
\label{Fig4}
\vspace{3cm}
\end{figure*}

Fig.~\ref{Fig4}A compares the experimentally obtained values of $N_{\rm sat}$ to their mean-field estimates in the BEC-BCS crossover, and the inset shows their ratios. The scaling factor $f$ has been determined to be $\approx 40$ by fitting the theory to the data using least-squares. The dependence of $N_{\rm sat}$ on $1/k_{\rm F}a$ is captured exceptionally well by the mean-field theory, where the slight variations of their ratios in the strongly interacting regime (shown in the inset) may imply the presence of beyond mean-field corrections to the healing length $\xi_{\rm h}$ and also possible weak variations in $f$. Fig.~\ref{Fig4}B summarizes the corresponding quench times $t_{\rm sat}$ where saturation begins to develop. Note that in order to account for the higher final trap depth used for the data taken at $-1/k_{\rm F}a=0.7$, the corresponding value of $t_{\rm sat}$ has been rescaled to match the quench rates. Interestingly, in contrast to $N_{\rm sat}$, $t_{\rm sat}$ maintains a fairly constant value within the investigated interaction strengths. Since the thermalization rate of the system would not show a sharp peak at unitarity~\cite{Gehm03}, a weak dependence of $t_{\rm sat}$ on $1/k_{\rm F}a$ is somewhat expected. However, a full understanding of this behavior will require an enhanced modeling of the quench profile at each interaction strength $1/k_{\rm F}a$ and the knowledge of the interaction dependence of the microscopic parameters $\xi_{0}$ and $\tau_{0}$ near the critical point, set by the thermal de-Broglie wavelength and the inverse thermalization rate of the system, respectively. Using the $N_{\rm sat}$ and $t_{\rm sat}$ extracted from the model fits to the data, the measured vortex numbers and their respective quench times can be rescaled to collapse the full data into a single universal curve representing the quench dynamics of vortex formation in this system throughout the different quench regimes, as shown in  Fig.~\ref{Fig4}C. This provides a benchmark for theoretical studies of quench production of defects in a strongly interacting Fermi superfluid and also for its related systems such as weakly interacting BECs and superfluid helium. 

Our measurements present a foray into the use of Feshbach resonances to precisely and broadly tune the role of interactions in the study of critical dynamics of a many-body system. This offers a highly consistent means of controlling the microscopic length and time scales that govern the system's critical dynamics. An extension of this work will be to perform a similar set of measurements in the weak-fluctuation limit, where the system's freeze-out correlation length is shorter than the Ginzburg length scale $\xi_{\rm G}$, and the mean-field description of its critical dynamics is restored. A precise measurement of the evolution of the KZ exponent as the system is tuned further away from unitarity in the BEC regime will characterize the role of fluctuations. Furthermore, the use of tailored optical potentials to confine the atoms in a quasi-2D geometry offers an additional means of controlling the role of fluctuations. In 2D, the Landau-Ginzburg description of phase transitions breaks down, and superfluidity arises through the infinite-order Berezinskii-Kosterlitz-Thouless (BKT) transition, whose exponential divergence of the correlation length and relaxation time has been predicted to give rise to a logarithmic correction to the KZ exponent~\cite{Jelic2011}.

\newpage

\bibliographystyle{Science}

\newpage

\section*{Acknowledgements}
This work was supported by the Institute for Basic Science in Korea (Grant No. IBS-R009-D1) and the National Research Foundation of Korea (Grant No. NRF-2018R1A2B3003373). JWP acknowledges support from the POSCO Science Fellowship of the POSCO TJ Park Foundation. The data supporting  this manuscript are available from the corresponding authors upon request.

\newpage
\setcounter{page}{1}
\beginsupplement
\section*{\centering{Supplementary Materials for \\\vspace{5mm}
Kibble-Zurek universality in a strongly interacting Fermi superfluid}}
\renewcommand\thesubsection{\arabic{subsection}}
\setcounter{secnumdepth}{2}
\setcounter{tocdepth}{2}
\renewcommand{\contentsname}{Materials and Methods}
\vspace{5mm}\tableofcontents\vspace{5mm}

\subsection{Sample preparation}

A detailed description of our experimental setup and the preparation of a strongly interacting Fermi gas of $^{6}$Li is given in Ref.~\cite{Park2018}. To create a large superfluid sample that can support a high number of vortices, we use bosonic $^{23}$Na to sympathetically cool fermionic $^{6}$Li to degeneracy. To this end, both atomic species are simultaneously loaded into a magneto-optical trap, optically pumped to their magnetically trappable stretched states, and subsequently transferred to a plugged magnetic quadrupole trap, where radio-frequency (rf) evaporation of $^{23}$Na cools the $^{6}$Li atoms to quantum degeneracy. Then, the sample is moved to a single-beam optical dipole trap (ODT) with an aspect ratio of 110:1 ($\lambda = 1064$ nm), and the remaining $^{23}$Na atoms are removed using a resonant light pulse. 

To access the broad $s$-wave Feshbach resonance between the two lowest hyperfine states of $^{6}$Li (denoted by $\lvert1\rangle = \lvert F{=}1/2, m_{F}{=}1/2 \rangle$ and $\lvert2\rangle = \lvert F{=}1/2, m_{F}{=}-1/2 \rangle$) located at 832 G, the $^{6}$Li atoms in the ODT are initially transferred to the $\lvert 1 \rangle$ state by an rf Landau-Zener sweep at 3 G, and the magnetic field is increased to 870 G, where another rf Landau-Zener sweep creates an equal mixture of $\lvert1\rangle$ and $\lvert2\rangle$. Subsequently, the magnetic field is ramped to 815 G on the BEC side of the resonance, where the sample is further cooled by reducing the ODT beam intensity to a value that corresponds to a trap depth of $U_{\rm i}=1.15~U_{\rm c}(B)$, where $U_{\rm c}(B)$ is the critical ODT depth for the onset of condensation at magnetic field $B$. Since $U_{\rm c}$ is dependent on the interaction strength, its value for a given atom number is determined at each $B$ accessed in the experiment by examining the condensate fraction as a function of the trap depth. Finally, the magnetic field is adiabatically ramped to $B$ where the thermal quench will be performed. 

At this stage of the experiment prior to the thermal quench, the gas is composed of approximately $2.0 {\times} 10^{6}$ $^{6}$Li atoms per spin state for all explored values of $B$. The trapping frequencies in the radial plane of the sample are $(\omega_{x}, \omega_{y}) = 2 \pi \times (20, 19)$ Hz, and the axial trapping frequency ranges between {$\omega_{z} = 2 \pi \times$ 1080 Hz at 757 G and 780 Hz at 898 G} depending on $B$. Here, the radially symmetric confinement is dominantly provided by the residual magnetic curvature of the Feshbach field, and the tight axial confinement is provided by the ODT. The variation of $\omega_{x}$ and $\omega_{y}$ for the explored range of $B$ is negligible.

\subsection{Thermal quenching}

\begin{figure*}[p]
\vspace{1cm}
\centering
\includegraphics[width=0.4\columnwidth]{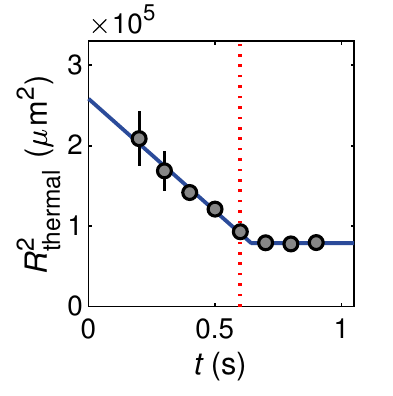}
\caption{The evolution of the square of the thermal cloud size in the experiment for 600 ms quench at $-1/k_{\rm F}a{=}-1.53$. Each data point comprises ten realizations of the same experiment, and the error bars represent the standard deviation. The blue line is a bilinear fit to the data, where one of the lines is kept horizontal. The vertical dotted line marks the end of the quench.
}
\label{RTF}
\vspace{3cm}
\end{figure*}

Once the sample is prepared, the strongly interacting Fermi gas is thermally quenched across the spontaneous symmetry-breaking normal to superfluid phase transition by linearly reducing the ODT trap beam intensity. From its initial value of $U_{\rm i}=1.15~U_{\rm c}(B)$, the intensity is ramped to a final value of $U_{\rm f} = 0.15~U_{\rm c}(B)$ ($U_{\rm f} = 0.3~U_{\rm c}$ for measurements at $B = 898$ G) in variable durations ranging between 0.2 s and 2.6 s. The linear relationship between the sample temperature and the ODT beam intensity can be inferred by observing the evolution of the thermal cloud size in time-of-flight images from the experiment. Fig.~\ref{RTF} shows that the square of the thermal cloud size, which is proportional to the sample temperature, linearly decreases as the intensity is linearly reduced during the quench, and it remains constant after the quench is completed.

\subsection{Detecting and counting vortices}

The spontaneously created vortices are manifested as density depleted holes in time-of-flight (TOF) images of the gas. The detection sequence begins by simultaneously switching off the ODT and rapidly decreasing the Feshbach field to a value close to zero. This initiates the TOF expansion of the sample and converts the fermion pairs into tightly bound molecules~\cite{Regal2004}. The sample freely expands for 13.5 ms, and then the magnetic field is quickly ramped up to 695 G in 10 ms, where an absorption image of the gas is taken. Due to the tight axial confinement provided by the ODT, the condensate rapidly expands in the axial direction upon its release, and the radial size remains fairly constant during the TOF.

For the experiments performed at 898 G on the BCS side of the resonance, the vortex visibility is reduced due to the higher thermal fraction of the sample. To enhance the depletion contrast and help identify the vortices, we ramp the magnetic field from 898 G to 855 G in 5 ms before releasing the gas from the trap. This additional ramp increases the condensate fraction of the sample and boosts the contrast of the depletion~\cite{Zwierlein2005}. We check that this procedure does not influence the number of observed vortices by comparing the number with and without the ramp.

\begin{figure*}[p]
\vspace{1cm}
\centering
\includegraphics[width=7cm]{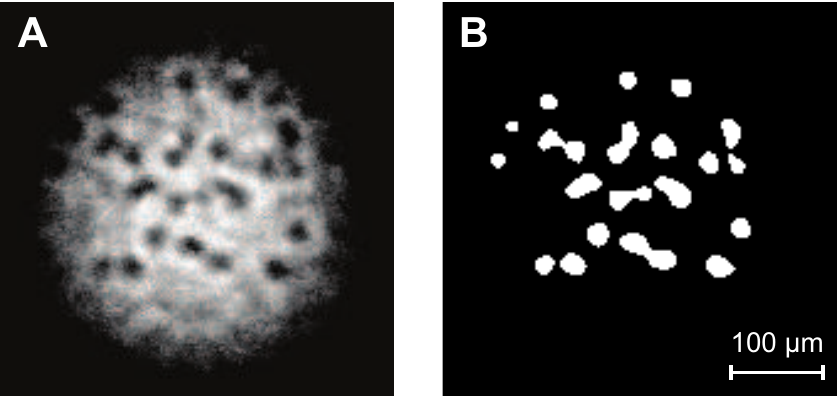}
\caption{Creating a binary image of the density depleted holes. ({\bf A}) is an exemplary image of the optical depth of the sample after TOF, and ({\bf B}) shows the converted binary image identifying the density depleted holes.}
\label{binary}
\vspace{3cm}
\end{figure*}

\begin{figure*}[p]
\vspace{1cm}
\centering
\includegraphics[width=1\columnwidth]{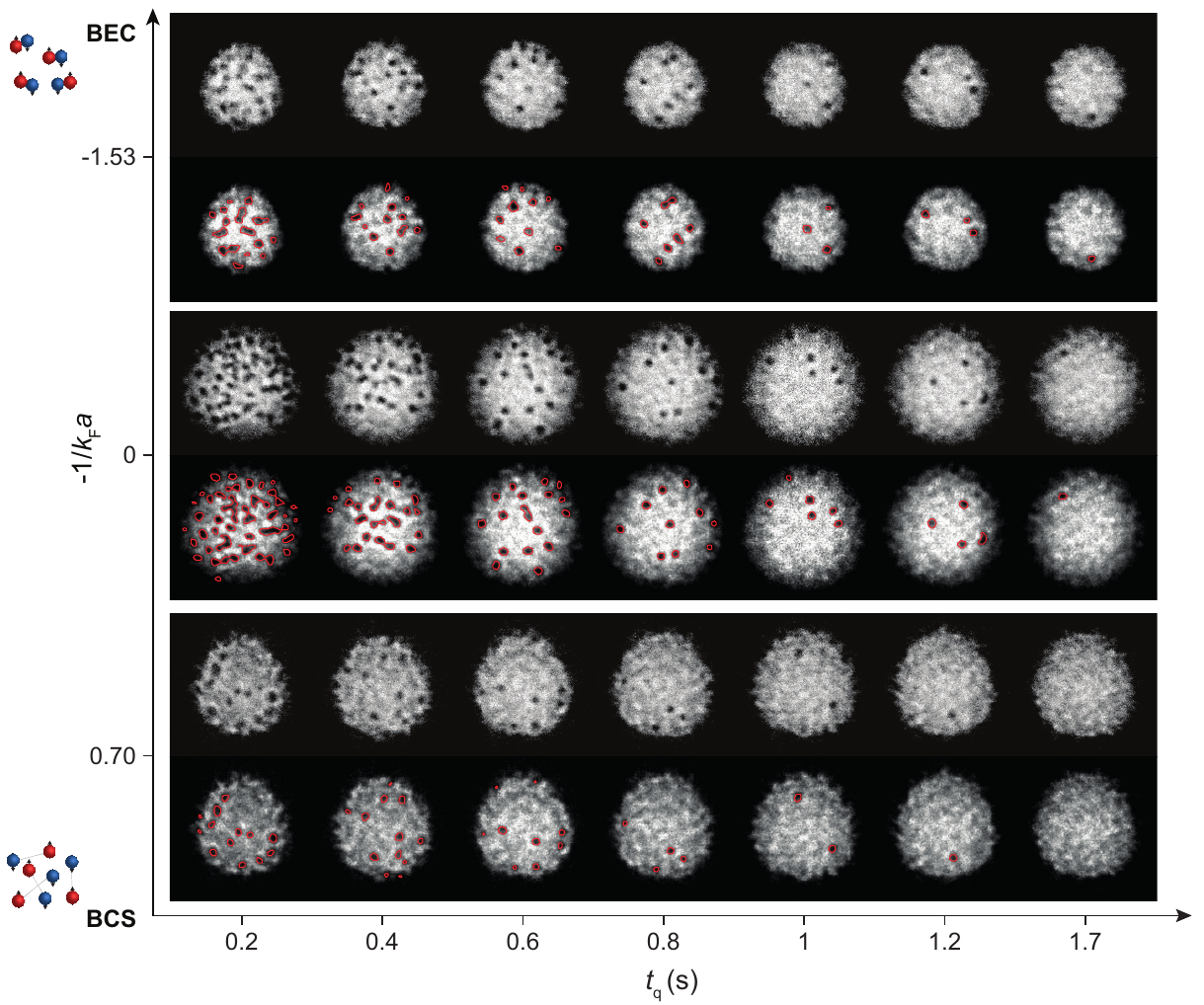}
\caption{Computer-assisted counting of vortex number. Images from Fig.~1D and when they are applied to the vortex counting procedure. Identified vortices are encompassed by red lines.}
\label{count}
\vspace{3cm}
\end{figure*}

To extract the vortex number from TOF images, we use an automized image processing method, similar to the one outlined in Ref.~\cite{Kwon2014}. First, for a given absorption image, the contribution to the optical depth from the non-condensed fraction of the sample is removed by fitting a Gaussian to the thermal wings and subtracting it off. Then, a copy of this image is created, and a two-dimensional Gaussian smoothing filter is applied, whose width is chosen to be comparable to the typical size of a density depleted hole. This filtered image is used to divide the unfiltered image, which is then binarized using an empirically chosen threshold value that best identifies the density depleted holes as isolated ``particles" [see Fig.~\ref{binary}]. Using a standard particle analysis package included in most mathematical computing software (e.g. Matlab, Igor, Mathematica), the boundaries and the areas of the individual particles are identified. Fig.~\ref{count} exhibits exemplary TOF images taken from Fig.~1D of the main manuscript together with the processed images showing the boundaries of the density depleted holes, demonstrating the reliability of this procedure.

\begin{figure*}[p]
\vspace{1cm}
\centering
\includegraphics[width=8.5cm]{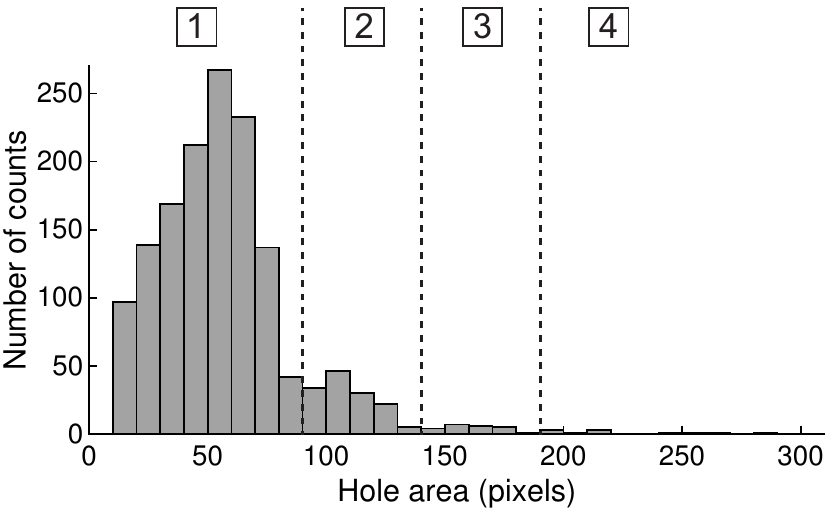}
\caption{The histogram of the identified vortex areas from 70 images at unitarity, where the quench time extends between 200 ms to 800 ms.}
\label{histo}
\vspace{3cm}
\end{figure*}

For samples that are densely populated by vortices, large holes that represent multiple vortices are observed. To assign a proper quanta, we plot the histogram of the hole area at each investigated interaction strength and assign a cut-off area for each quanta based on the multiple peak structure of the histogram [see Fig.~\ref{histo}]. The minima between two adjacent peaks are used to set the vortex number transition lines of the particles. Based on this criteria, each particle is assigned a quanta equal to the number of vortices it represents, and their sum is recorded as the number of vortices of the image. Also, once the vortex number is determined by this procedure, every image was double-checked by eye to correct for possible misassignments.

\subsection{Condensate formation and vortex decay}
 
In the KZ mechanism, topological defects emerge in the system through the merging of domains with independent order parameters. To reliably extract the number of defects, a certain amount of hold time must be applied to the sample after the thermal quench to ensure that the condensate growth and the domain merging dynamics have been completed. However, in the presence of destructive interactions among the defects, this hold time must be kept as short as possible, such that the ensuing reduction of their numbers does not affect the observed scaling relationship between the defect density and the quench rate.

\begin{figure*}[p]
\vspace{1cm}
\centering
\includegraphics[width=1\columnwidth]{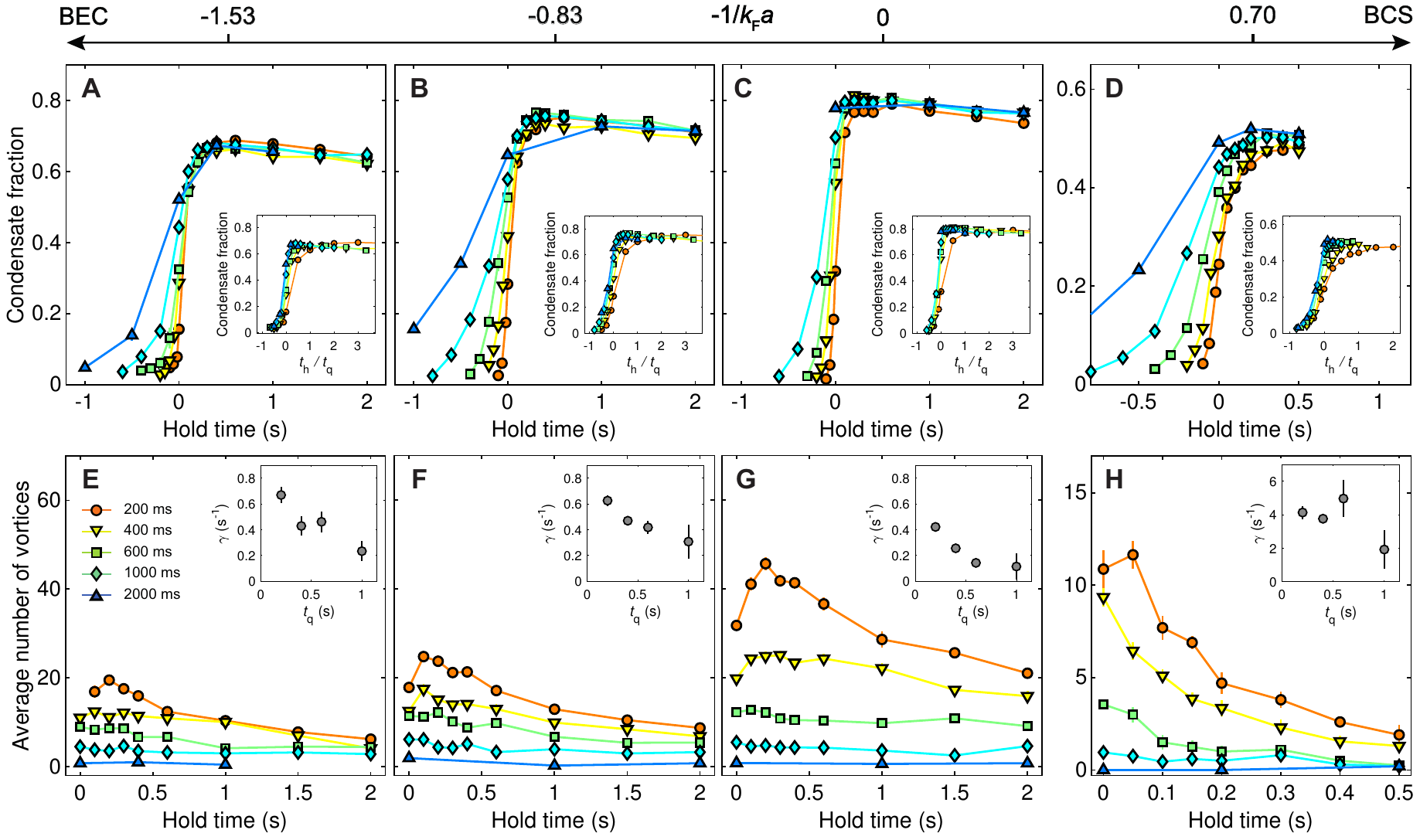}
\caption{The evolution of the sample during and after various quench times $t_{\rm q}=200$ ms (circle), 400 ms (inverted triangle), 600 ms (square), 1000 ms (diamond), and 2000 ms (triangle). ({\bf A})-({\bf D}) The growth and decay of the condensate fraction of the sample during and after the quench. ({\bf E})-({\bf H}) The decay of the average number of counted vortices during hold time $t_{\rm h}$ after the quench. Each data point is the average of ten realizations of the same experiment and the error bars give the standard error of the mean.}
\label{decay}
\vspace{3cm}
\end{figure*}

It should be noted that for the case of weakly interacting BECs, investigations on the effect of the dissipative evolution of the defects on the observed KZ scaling have shown that the KZ exponent is fairly robust against the decay of the defect number~\cite{Donadello2016, Liu2018}. Nonetheless, to apply a well-defined hold time $t_{\rm h}$ to the sample before releasing it for TOF imaging, we investigate the evolution of the condensate fraction and the vortex number as a function of $t_{\rm h}$ for a number of quench times, at each the interaction strength accessed in the experiment [see Fig.~\ref{decay}].

Generically, following the quench, the condensate fraction initially rises until it reaches a maximal value near $t_{\rm h} = 200 \sim 300$ ms ($50 \sim 150$ ms for measurements on the BCS side), and then it decays due to three-body losses. This value of $t_{\rm h}$ may represent the intrinsic time scale for the condensate growth dynamics of the system. The condensate fraction measured after this hold time is independent of the quench time for each $-1/k_{\rm F}a$, indicating that the sample is well thermalized during the quench for the investigated quench times. Also, when the hold time of each data set is normalized by its respective quench time [see insets of Fig.~\ref{decay}A-D], the evolution of the condensate fraction during the ODT ramp closely tracks each other for all the quench times apart from 200 ms, where it starts to show a weak deviation. This observation implies that the temperature of the sample is set by the ODT depth for the explored quench times during the quench.

The evolution of the vortex number shows a similar trend compared to that of the condensate fraction. Specifically, the area of the density depleted regions initially rises as the domains merge until it reaches a maximal value, and then it decays as the loss mechanism among the defects starts to dominate. Based on these observations, we set $t_{\rm h}$ to be the time at which the defect number reaches its maximum, corresponding to $t_{\rm h} = 200$ ms for measurements performed on the BEC side of the resonance and at unitarity, and $t_{\rm h} = 50$ ms for the data taken on the BCS side due to a faster decay rate. 

An important observation is that the decay rate of the vortices is dependent on the initial vortex density set by the quench time $t_{\rm q}$. Specifically, it becomes enhanced at short quench times where the initial vortex number is higher. The insets in Fig.~\ref{decay}E-H show the decay constant $\gamma$ as a function of the quench time when we fit an exponential decay of the vortex number $N(t) = N_{0}e^{-\gamma t}$ to the data whose hold time is equal to or greater than $t_{\rm h}$. Here, $t$ is the hold time and $N_0$ is the the hypothetical vortex number at $t=0$. The increase in the exponential decay rate for shorter quench times reveals the presence of a beyond one-body decay mechanism, which likely arises from the destructive interactions among vortices with opposite charges. In the experiment, this loss mechanism is associated with the departure from the KZ scaling and the saturation of the vortex number for short quench times.

\subsection{Mean-field critical velocity}

The mean-field Landau critical velocity of a strongly interacting Fermi superfluid in the BEC-BCS crossover is given by min$(v_{\rm s}, v_{\rm pb})$ where $v_{\rm s}$ is the speed of sound and $v_{\rm pb}$ is the mean-field BCS pair-breaking velocity of the superfluid~\cite{Combescot2006}. 
The speed of sound is obtained from the quantum Monte Carlo equation of state in Ref.~\cite{Manini2005}.
For our inhomogeneous superfluid sample in the harmonic potential, the critical velocity $v_{\rm c}$ at its center is calculated by assuming the local density approximation. 
Here, the column averaged density, instead of the central density, has to be employed in computing $v_{\rm s}$ since the superfluid is hydrodynamic in the axial direction.

\newpage

\newpage

\end{document}